\documentclass[conference]{IEEEtran}
\IEEEoverridecommandlockouts
\usepackage{cite}
\usepackage{amsmath,amssymb,amsfonts}
\usepackage{graphicx}
\usepackage{textcomp}
\usepackage{xcolor}

\usepackage{tikz}

\usepackage{subfig}

\usepackage{physics}

\usepackage{algorithm, algpseudocode} 

\usepackage{caption,setspace}
\def\BibTeX{{\rm B\kern-.05em{\sc i\kern-.025em b}\kern-.08em
    T\kern-.1667em\lower.7ex\hbox{E}\kern-.125emX}}
\begin{document}

\title{A Configurable Protocol for Quantum Entanglement Distribution to End Nodes\\
\thanks{}
}

\author{\IEEEauthorblockN{Leonardo Bacciottini}
\IEEEauthorblockA{\textit{Dept. of Information Engineering} \\
\textit{Universities of Florence and Pisa}\\
Florence, Pisa, Italy \\
leonardo.bacciottini@phd.unipi.it}
\and
\IEEEauthorblockN{Luciano Lenzini}
\IEEEauthorblockA{\textit{Dept. of Information Engineering} \\
\textit{University of Pisa}\\
Pisa, Italy \\
lenzini44@gmail.com}
\and
\IEEEauthorblockN{Enzo Mingozzi}
\IEEEauthorblockA{\textit{Dept. of Information Engineering} \\
\textit{University of Pisa}\\
Pisa, Italy \\
enzo.mingozzi@unipi.it}
\and
\IEEEauthorblockN{Giuseppe Anastasi}
\IEEEauthorblockA{\textit{Dept. of Information Engineering} \\
\textit{University of Pisa}\\
Pisa, Italy \\
giuseppe.anastasi@unipi.it}
}

\author{\IEEEauthorblockN{Leonardo Bacciottini}
\IEEEauthorblockA{\textit{Dept. of Information Engineering} \\
\textit{Universities of Florence and Pisa}\\
Florence and Pisa, Italy\\
leonardo.bacciottini@phd.unipi.it}
\and
\IEEEauthorblockN{Luciano Lenzini}
\IEEEauthorblockA{\textit{Dept. of Information Eng.} \\
\textit{University of Pisa}\\
Pisa, Italy\\
lenzini44@gmail.com}
\and
\IEEEauthorblockN{Enzo Mingozzi}
\IEEEauthorblockA{\textit{Dept. of Information Eng.} \\
\textit{University of Pisa}\\
Pisa, Italy\\
enzo.mingozzi@unipi.it}
\and
\IEEEauthorblockN{Giuseppe Anastasi}
\IEEEauthorblockA{\textit{Dept. of Information Eng.} \\
\textit{University of Pisa}\\
Pisa, Italy\\
giuseppe.anastasi@unipi.it}
}
\maketitle

\begin{tikzpicture}[remember picture,overlay] \node[anchor=south,yshift=10pt] at (current page.south) {\fbox{\parbox{\dimexpr\textwidth-\fboxsep-\fboxrule\relax}{
  \footnotesize{
    © 2023 IEEE.  Personal use of this material is permitted.  Permission from IEEE must be obtained for all other uses, in any current or future media, including reprinting/republishing this material for advertising or promotional purposes, creating new collective works, for resale or redistribution to servers or lists, or reuse of any copyrighted component of this work in other works.
  }
}}};
\end{tikzpicture}

\begin{abstract}
The primary task of a quantum repeater network is to deliver entanglement among end nodes. Most of existing entanglement distribution protocols do not consider purification, which is thus delegated to an upper layer. This is a major drawback since, once an end-to-end entangled connection (or a portion thereof) is established it cannot be purified if its fidelity (F) does not fall within an interval bounded by Fmin (greater than 0.5) and Fmax (less than 1). In this paper, we propose the Ranked Entanglement Distribution Protocol (REDiP), a connection-oriented protocol that overcomes the above drawback. This result was achieved by including in our protocol two mechanisms for carrying out jointly purification and entanglement swapping. We use simulations to investigate the impact of these mechanisms on the performance of a repeater network, in terms of throughput and fidelity. Moreover, we show how REDiP can easily be configured to implement custom entanglement swapping and purification strategies, including (but not restricted to) those adopted in two recent works.
\end{abstract}


\enlargethispage{16pt}

\section{Introduction}
In the last few years there have been different proposals for a quantum Internet protocol stack \cite{Illiano_2022}, but a common trait of all near-term quantum network architectures is entanglement distribution through the \emph{Purify-and-Swap} scheme \cite{Dur_1999}, which glues many entanglement connections between adjacent quantum repeaters into an end-to-end entangled pair. This will probably change in the future when more advanced quantum hardware will enable support for quantum error correcting codes \cite{Jiang_2009}. This work focuses on the former approach, as it can be achieved with near-term hardware.

\begin{figure} [tb]
    \captionsetup[subfloat]{farskip=5pt,captionskip=3pt}
    \centering
    \subfloat[]{\includegraphics[width=.9\linewidth]{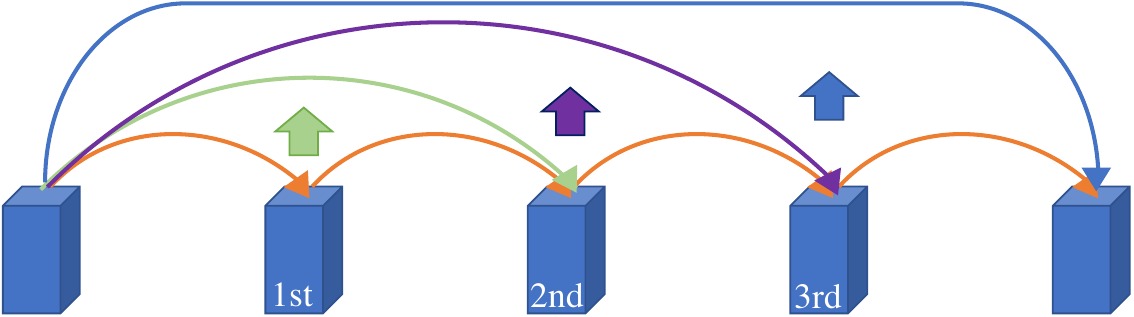}\label{fig:consecutive_ess}}
    \hfill
    \subfloat[]{\includegraphics[width=.9\linewidth]{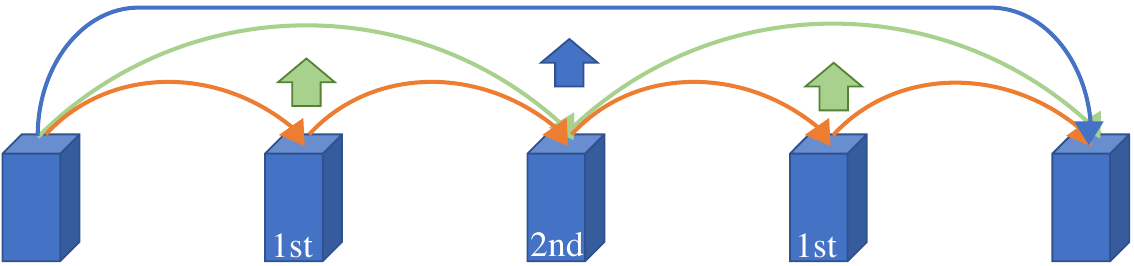}\label{fig:nested_ess}}
    \hfill
    \subfloat[]{\includegraphics[width=.9\linewidth]{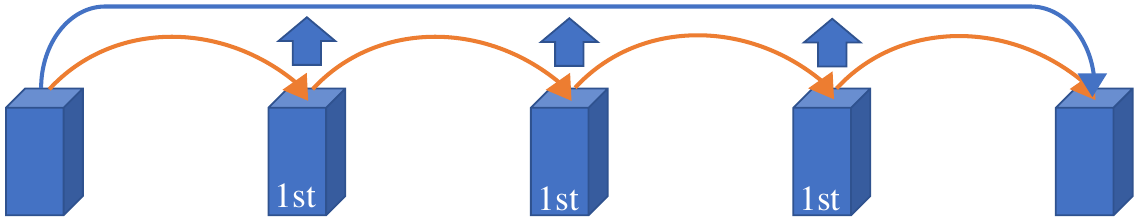}\label{fig:parallel_ess}}
    \caption{Three entanglement swapping strategies on a five nodes chain (two end nodes and three repeaters): (a) Consecutive, (b) Nested, (c) Parallel. The arcs represent entanglement between the two connected nodes, whereas the arrows and the legend specify the entanglement swapping order.}
    \label{fig:ess_strategies}
\end{figure}

To be completed, entanglement swapping also requires the transmission of two bits of classical information to let one of the recipients know in which Bell state the qubit pair has landed. This requirement introduces a new problem, that is determining the order in which this classical synchronization between repeaters takes place. Such decision is very important because it defines – and is defined by - the order of performing entanglement swapping among nodes, as shown in Fig. \ref{fig:ess_strategies}. We call this choice \emph{Entanglement Swapping Strategy} (ESS). In this paper we focus on three different ESSs: (i) Consecutive ESS (Fig. \ref{fig:consecutive_ess}), where a node performs entanglement swapping only after receiving the classical message from the previous node on the chain, and then sends a message to the next node. (ii) Nested ESS (Fig. \ref{fig:nested_ess}), where the swapping order is recursively determined by extracting the nodes of the chain in odd positions - with positions starting from zero - until only the end nodes remain. (iii) Parallel ESS (Fig. \ref{fig:parallel_ess}), where all intermediate nodes perform entanglement swapping concurrently and transmit the result directly to one of the end nodes. If we include entanglement purification in this discussion, we define the \emph{Purification and  Entanglement Swapping Strategy} (PESS) as the policy that determines both the order of entanglement swapping other than when and where (i.e., at what node) purification rounds are to be carried out.

To quote only two proposals on several, in 2020 Kozlowski, Dahlberg and Wehner \cite{Kozlowski_2020} defined a connection-oriented protocol for entanglement distribution that exploits Parallel ESS to counter the effect of low memory coherence times. Moreover, in 2022 Li, Xue, Wei, and Yu \cite{Li_2022} designed another connection-oriented protocol with a detailed description of the connection establishment phase, involving a resource allocation mechanism based on quantum memory slots partitioning. Their protocol uses Consecutive ESS to have a strict control over the timing of the actions performed by each node. The limit of these previous works is the fact that these protocols are tightly bonded to a specific ESS and do not support entanglement purification as an integrated mechanism. This leads to a lack of flexibility that is proved in section III. Of course, it is always possible to have a recursive protocol stack where several connections and purification protocols are installed one on top of the other to implement an arbitrary PESS, as done in \cite{Aparicio_2011} and suggested in \cite{Kozlowski_2020}, but this inevitably introduces an overhead and complicates network configuration and resource allocation.

In this paper we present the \emph{Ranked Entanglement Distribution Protocol} (REDiP). Our protocol overcomes the aforementioned limits by letting users configure each single connection to implement the ESS that better suits their needs. We point out that REDiP is not stuck to the above strategies only. It also allows the implementation of any user-defined ESS. Moreover, the protocol empowers ESSs by enabling them to cooperate with entanglement purification to adopt an arbitrary PESS. Our results show that this configurability is essential to satisfy user requirements in a quantum network subject to quantum errors and link conditions that may change over time and space.

This paper is structured as follows: section \ref{sec:2} describes REDiP design, then section \ref{sec:3} shows the results of a simulation campaign that evaluates the performances of different PESSs implemented on REDiP under varying scenarios. Finally, section \ref{sec:4} draws some final remarks and future work.

\section {Protocol Design}
\label{sec:2}
REDiP is a connection-oriented protocol. Therefore, a connection on a path between the two end nodes must be created before distributing the entanglement. This connection conveys classical bits and is provided by a conventional infrastructure (for example, the classical Internet). To avoid ambiguities with entangled connections, we will refer to this preliminary connection as \emph{tunnel}. Since the tunnel establishment is not the main focus of this paper, we refer to the work done in [4] and [5] as to how this phase can be carried out with a one-way or two-way handshake respectively. This approach is borrowed from classical Internet protocols like the Resource Reservation Protocol (RSVP).

On the other hand, taking as reference the quantum protocol stack introduced in \cite{Dahlberg_2019}, REDiP is placed at the \emph{Network layer}. The underlying \emph{Link layer} has the task of generating heralded entangled Bell pairs on physical links that are consumed by Network layer protocols to distribute end-to-end entanglement. We will simply refer to upper layer protocols as \emph{Users} of REDiP. Before delving deeper into the analysis of the protocol, we define some reference variables that will be used across this section:
\begin{itemize}
    \item $N$ is the number of nodes composing the path.
    \item $L$ is the vector of node identifiers, in the order they are traversed. $L_i$ indicates the $ith$ node on the path, $i\in\{0,\ldots,N-1\}$.
    \item $R$ is the vector of node ranks, where each element $R_i$ is an integer indicating the rank of node $L_i$ (see section \ref{subsec:2a}).
    \item $P$ is the purification vector, where each element $P_r$ is an integer indicating how many purification rounds must be performed by nodes whose rank is $r$, $r\in\{0,\ldots,max\{R_i\in R\}\}$ (see section \ref{subsec:2a}).
    \item $K$ is the number of end-to-end pairs that a given tunnel must deliver to the destination end nodes.
\end{itemize}
Finally, we assume that the quantum network architecture where REDiP is used has an addressing scheme where each entangled qubit pair has an identifier shared at least among the nodes holding the two ends of the pair. For the sake of clarity, we call an entangled qubit pair between adjacent nodes an \emph{entangled link} or just a \emph{link}. We also use the term \emph{entangled segment} or just \emph{segment} when the entangled pair resides on non-adjacent nodes. Finally, we call \emph{end-to-end entangled connection} or just \emph{connection} an entangled segment that resides on the end nodes. Clearly the two qubits of an entangled pair embody the \emph{endpoints} of a link, segment, or connection.

\subsection{Opening and Configuring the Tunnel}
\label{subsec:2a}
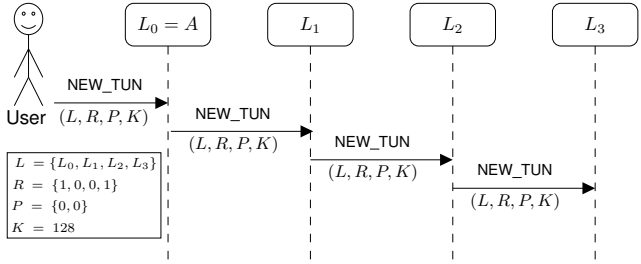
\begin{figure} [tb]
    \centering
    {\fontfamily{phv}\selectfont
            \resizebox{.96\linewidth}{!}{%
                \begin{tikzpicture}[x=0.75pt,y=0.75pt,yscale=-1,xscale=1]

\draw   (67.04,155) -- (175,155) -- (175,215) -- (67.04,215) -- cycle ;
\draw   (150,56) .. controls (150,52.69) and (152.69,50) .. (156,50) -- (204,50) .. controls (207.31,50) and (210,52.69) .. (210,56) -- (210,74) .. controls (210,77.31) and (207.31,80) .. (204,80) -- (156,80) .. controls (152.69,80) and (150,77.31) .. (150,74) -- cycle ;

\draw   (250,56) .. controls (250,52.69) and (252.69,50) .. (256,50) -- (304,50) .. controls (307.31,50) and (310,52.69) .. (310,56) -- (310,74) .. controls (310,77.31) and (307.31,80) .. (304,80) -- (256,80) .. controls (252.69,80) and (250,77.31) .. (250,74) -- cycle ;
\draw   (350,56) .. controls (350,52.69) and (352.69,50) .. (356,50) -- (404,50) .. controls (407.31,50) and (410,52.69) .. (410,56) -- (410,74) .. controls (410,77.31) and (407.31,80) .. (404,80) -- (356,80) .. controls (352.69,80) and (350,77.31) .. (350,74) -- cycle ;
\draw   (450,56) .. controls (450,52.69) and (452.69,50) .. (456,50) -- (504,50) .. controls (507.31,50) and (510,52.69) .. (510,56) -- (510,74) .. controls (510,77.31) and (507.31,80) .. (504,80) -- (456,80) .. controls (452.69,80) and (450,77.31) .. (450,74) -- cycle ;
\draw   (72,61) .. controls (72,54.92) and (76.48,50) .. (82,50) .. controls (87.52,50) and (92,54.92) .. (92,61) .. controls (92,67.08) and (87.52,72) .. (82,72) .. controls (76.48,72) and (72,67.08) .. (72,61) -- cycle ; \draw   (77.6,57.26) .. controls (77.6,56.65) and (78.05,56.16) .. (78.6,56.16) .. controls (79.15,56.16) and (79.6,56.65) .. (79.6,57.26) .. controls (79.6,57.87) and (79.15,58.36) .. (78.6,58.36) .. controls (78.05,58.36) and (77.6,57.87) .. (77.6,57.26) -- cycle ; \draw   (84.4,57.26) .. controls (84.4,56.65) and (84.85,56.16) .. (85.4,56.16) .. controls (85.95,56.16) and (86.4,56.65) .. (86.4,57.26) .. controls (86.4,57.87) and (85.95,58.36) .. (85.4,58.36) .. controls (84.85,58.36) and (84.4,57.87) .. (84.4,57.26) -- cycle ; \draw   (77,65.4) .. controls (80.33,68.33) and (83.67,68.33) .. (87,65.4) ;
\draw    (82,72) -- (82,102) ;
\draw    (82,102) -- (72,122) ;
\draw    (82,102) -- (92,122) ;
\draw    (82,82) -- (72,102) ;
\draw    (82,82) -- (92,102) ;

\draw    (100,120) -- (177,120) ;
\draw [shift={(180,120)}, rotate = 180] [fill={rgb, 255:red, 0; green, 0; blue, 0 }  ][line width=0.08]  [draw opacity=0] (8.93,-4.29) -- (0,0) -- (8.93,4.29) -- cycle    ;
\draw    (182,140) -- (279,140) ;
\draw [shift={(282,140)}, rotate = 180] [fill={rgb, 255:red, 0; green, 0; blue, 0 }  ][line width=0.08]  [draw opacity=0] (8.93,-4.29) -- (0,0) -- (8.93,4.29) -- cycle    ;

\draw  [dash pattern={on 4.5pt off 4.5pt}]  (180,80) -- (180,230) ;
\draw  [dash pattern={on 4.5pt off 4.5pt}]  (280,80) -- (280,230) ;
\draw  [dash pattern={on 4.5pt off 4.5pt}]  (380,80) -- (380,230) ;
\draw  [dash pattern={on 4.5pt off 4.5pt}]  (480,80) -- (480,230) ;
\draw    (280,160) -- (377,160) ;
\draw [shift={(380,160)}, rotate = 180] [fill={rgb, 255:red, 0; green, 0; blue, 0 }  ][line width=0.08]  [draw opacity=0] (8.93,-4.29) -- (0,0) -- (8.93,4.29) -- cycle    ;

\draw    (380,180) -- (477,180) ;
\draw [shift={(480,180)}, rotate = 180] [fill={rgb, 255:red, 0; green, 0; blue, 0 }  ][line width=0.08]  [draw opacity=0] (8.93,-4.29) -- (0,0) -- (8.93,4.29) -- cycle    ;

\draw (180,65) node   [align=left] {$\displaystyle L_{0} =A$};
\draw (65,124) node [anchor=north west][inner sep=0.75pt]   [align=left] {User};
\draw (108,102) node [anchor=north west][inner sep=0.75pt]  [font=\footnotesize] [align=left] {NEW\_TUN};
\draw (280,65) node   [align=left] {$\displaystyle L_{1}$};
\draw (380,65) node   [align=left] {$\displaystyle L_{2}$};
\draw (480,65) node   [align=left] {$\displaystyle L_{3}$};
\draw (102,123) node [anchor=north west][inner sep=0.75pt]  [font=\small] [align=left] {$\displaystyle ( L,R,P,K)$};
\draw (193,142) node [anchor=north west][inner sep=0.75pt]  [font=\small] [align=left] {$\displaystyle ( L,R,P,K)$};
\draw (71.05,157) node [anchor=north west][inner sep=0.75pt]  [font=\scriptsize] [align=left] {$\displaystyle L\ =\{L_{0} ,L_{1} ,L_{2} ,L_{3}\}$};
\draw (69.43,172) node [anchor=north west][inner sep=0.75pt]  [font=\scriptsize] [align=left] {$\displaystyle R\ =\ \{1,0,0,1\}$};
\draw (68.76,187) node [anchor=north west][inner sep=0.75pt]  [font=\scriptsize] [align=left] {$\displaystyle P\ =\ \{0,0\}$};
\draw (68.53,203) node [anchor=north west][inner sep=0.75pt]  [font=\scriptsize] [align=left] {$\displaystyle K\ =\ 128$};
\draw (198,122) node [anchor=north west][inner sep=0.75pt]  [font=\footnotesize] [align=left] {NEW\_TUN};
\draw (296,142) node [anchor=north west][inner sep=0.75pt]  [font=\footnotesize] [align=left] {NEW\_TUN};
\draw (291,162) node [anchor=north west][inner sep=0.75pt]  [font=\small] [align=left] {$\displaystyle ( L,R,P,K)$};
\draw (396,162) node [anchor=north west][inner sep=0.75pt]  [font=\footnotesize] [align=left] {NEW\_TUN};
\draw (391,182) node [anchor=north west][inner sep=0.75pt]  [font=\small] [align=left] {$\displaystyle ( L,R,P,K)$};

\end{tikzpicture}

            }
    }
    \vspace{-.0cm}
    \caption{A simplified example of REDiP tunnel establishment.}
    \label{fig:3}
\end{figure}

REDiP becomes active at an end node $A$ when a user, typically an upper layer protocol, submits a NEW\_TUN message containing the four $(L,R,P,K)$ variables. As shown in Fig. \ref{fig:3}, this triggers the RSVP-like tunnel establishment where resources are reserved and the tuple $(L,R,P,K)$ is shared among all nodes along the path. Furthermore, as in \cite{Kozlowski_2020}, we assume that the Link layer protocol continuously generates entangled links once the tunnel is open. For any given node, from now on we will refer to the directions towards the rightmost and leftmost nodes on the path $L$ as \emph{upstream} and \emph{downstream} respectively.

Entanglement swapping is carried out in consecutive steps. The rank of a node, specified inside the vector $R$, indicates the step at which that node must swap. By design, nodes whose rank is zero will swap entangled links as soon as they can, whereas nodes whose rank is $r>0$ are allowed to swap an entangled segment only after a (classical) signal from an $(r-1)$-ranked node notifies that the segment is ready. A constraint for REDiP to work properly is that end nodes of the connection must share the same maximum rank $r_{max}$, so that they are notified only when the end-to-end connection is ready. For what concerns purification, the user can specify how many purification rounds should be performed by nodes belonging to each specific rank. For example, a vector $P=[2,0,1]$ means that $0$-ranked nodes purify entangled links twice before swapping, 1-ranked nodes do not need purification, whereas $2$-ranked nodes (end nodes in this example) purify again once before delivering the end-to-end connection to the user. REDiP allows the negotiation of any purification algorithm supported by the nodes’ hardware during the tunnel establishment phase as part of the resource allocation process. Once established, the tunnel remains open until $K$ end-to-end connections have been delivered and consumed by the user.

\subsection{Generating Entangled Connections}
\label{subsec:2b}
REDiP uses the allocated tunnel to generate entangled connections. The control operations of this phase are carried out by three types of classical message called (i) SWAP\_UPDATE, used to notify nodes about the result of an entanglement swapping, (ii) PURIF\_SOLICIT and (iii) PURIF\_RESPONSE used during the entanglement purification procedure. We will gradually go through their definition and usage as the description moves forward.

When the tunnel has been established, each intermediate node $L_i$ processes the parameters $(L,R,P,K)$ and computes the following variables:
\begin{itemize}
    \item 	$L_i$ Swapping Destinations ($SD^i$), a two components vector where its components ${SD}_0^i$ and ${SD}_1^i$ are picked as the downstream and upstream nodes closest to $L_i$ whose rank is higher than $R_i$,
    \item 	$L_i$ Swapping Neighbors ($SN^i$), a two components vector where its components ${SN}_0^i$ and ${SN}_1^i$ are picked as the downstream and upstream nodes closest to $L_i$, whose rank is \emph{equal or} higher than $R_i$.
\end{itemize}
These variables are used to determine whether an intermediate node $L_i$ should generate or not a SWAP\_UPDATE message after the execution of the entanglement swapping. Taking the Nested ESS (Fig. \ref{fig:nested_ess}) as an example, all nodes must generate both an upstream and a downstream SWAP\_UPDATE after they swap and send them to their swapping destinations. On the contrary, in the Parallel ESS (Fig. \ref{fig:parallel_ess}) only the second and the penultimate nodes on the path generate an upstream and downstream SWAP\_UPDATE respectively. All other nodes limit to update and forward incoming updates to their swapping neighbors.  To implement this behavior, the node $L_i$  generates one or two SWAP\_UPDATEs only if ${SD}^i$ and ${SN}^i$ partially or totally coincide. We provide a more specific description in section \ref{subsubsec:updates}.

\begin{figure} [tb]
    \centering
    {\fontfamily{phv}\selectfont
            \resizebox{\linewidth}{!}{%
                \begin{tikzpicture}[x=0.75pt,y=0.75pt,yscale=-1,xscale=1]

\draw  [color={rgb, 255:red, 97; green, 50; blue, 50 }  ,draw opacity=1 ] (61,110) .. controls (61,90.67) and (76.67,75) .. (96,75) .. controls (115.33,75) and (131,90.67) .. (131,110) .. controls (131,129.33) and (115.33,145) .. (96,145) .. controls (76.67,145) and (61,129.33) .. (61,110) -- cycle ;

\draw   (300,109) .. controls (300,89.67) and (315.67,74) .. (335,74) .. controls (354.33,74) and (370,89.67) .. (370,109) .. controls (370,128.33) and (354.33,144) .. (335,144) .. controls (315.67,144) and (300,128.33) .. (300,109) -- cycle ;

\draw   (459,309) .. controls (459,289.67) and (474.67,274) .. (494,274) .. controls (513.33,274) and (529,289.67) .. (529,309) .. controls (529,328.33) and (513.33,344) .. (494,344) .. controls (474.67,344) and (459,328.33) .. (459,309) -- cycle ;

\draw   (140.5,259) .. controls (140.5,239.67) and (156.17,224) .. (175.5,224) .. controls (194.83,224) and (210.5,239.67) .. (210.5,259) .. controls (210.5,278.33) and (194.83,294) .. (175.5,294) .. controls (156.17,294) and (140.5,278.33) .. (140.5,259) -- cycle ;

\draw    (118,83) .. controls (148.54,49.51) and (267.36,46.1) .. (308.19,82.32) ;
\draw [shift={(310,84)}, rotate = 224.26] [fill={rgb, 255:red, 0; green, 0; blue, 0 }  ][line width=0.08]  [draw opacity=0] (8.93,-4.29) -- (0,0) -- (8.93,4.29) -- cycle    ;
\draw   (541,109) .. controls (541,89.67) and (556.67,74) .. (576,74) .. controls (595.33,74) and (611,89.67) .. (611,109) .. controls (611,128.33) and (595.33,144) .. (576,144) .. controls (556.67,144) and (541,128.33) .. (541,109) -- cycle ;

\draw    (360,84) .. controls (390.54,50.51) and (509.36,47.1) .. (550.19,83.32) ;
\draw [shift={(552,85)}, rotate = 224.26] [fill={rgb, 255:red, 0; green, 0; blue, 0 }  ][line width=0.08]  [draw opacity=0] (8.93,-4.29) -- (0,0) -- (8.93,4.29) -- cycle    ;
\draw    (362.47,136.76) .. controls (404.06,180.82) and (503,176.16) .. (552,135) ;
\draw [shift={(360,134)}, rotate = 49.6] [fill={rgb, 255:red, 0; green, 0; blue, 0 }  ][line width=0.08]  [draw opacity=0] (8.93,-4.29) -- (0,0) -- (8.93,4.29) -- cycle    ;
\draw    (532.54,291.2) .. controls (573.48,245.7) and (585.1,213.58) .. (590,144) ;
\draw [shift={(530,294)}, rotate = 312.46] [fill={rgb, 255:red, 0; green, 0; blue, 0 }  ][line width=0.08]  [draw opacity=0] (8.93,-4.29) -- (0,0) -- (8.93,4.29) -- cycle    ;
\draw    (456.78,299.06) .. controls (351.29,300.16) and (316.29,209.33) .. (345,143) ;
\draw [shift={(460,299)}, rotate = 178.41] [fill={rgb, 255:red, 0; green, 0; blue, 0 }  ][line width=0.08]  [draw opacity=0] (8.93,-4.29) -- (0,0) -- (8.93,4.29) -- cycle    ;
\draw    (201.91,230.74) .. controls (232.9,178.2) and (255.24,158.62) .. (316,140) ;
\draw [shift={(200,234)}, rotate = 300.19] [fill={rgb, 255:red, 0; green, 0; blue, 0 }  ][line width=0.08]  [draw opacity=0] (8.93,-4.29) -- (0,0) -- (8.93,4.29) -- cycle    ;
\draw    (203,282) .. controls (256.46,354.27) and (396.17,364.79) .. (472.7,339.77) ;
\draw [shift={(475,339)}, rotate = 161.11] [fill={rgb, 255:red, 0; green, 0; blue, 0 }  ][line width=0.08]  [draw opacity=0] (8.93,-4.29) -- (0,0) -- (8.93,4.29) -- cycle    ;
\draw  [dash pattern={on 0.84pt off 2.51pt}]  (85,222) .. controls (139.88,203.38) and (138.1,180.92) .. (119.18,141.44) ;
\draw [shift={(118,139)}, rotate = 64] [fill={rgb, 255:red, 0; green, 0; blue, 0 }  ][line width=0.08]  [draw opacity=0] (8.93,-4.29) -- (0,0) -- (8.93,4.29) -- cycle    ;
\draw  [dash pattern={on 0.84pt off 2.51pt}]  (526,325) .. controls (565.4,321.06) and (609.65,321.97) .. (627.22,370.74) ;
\draw [shift={(628,373)}, rotate = 251.57] [fill={rgb, 255:red, 0; green, 0; blue, 0 }  ][line width=0.08]  [draw opacity=0] (8.93,-4.29) -- (0,0) -- (8.93,4.29) -- cycle    ;
\draw  [dash pattern={on 0.84pt off 2.51pt}]  (155,286) .. controls (149.12,349.7) and (94.26,339.45) .. (73.24,300.42) ;
\draw [shift={(72,298)}, rotate = 64] [fill={rgb, 255:red, 0; green, 0; blue, 0 }  ][line width=0.08]  [draw opacity=0] (8.93,-4.29) -- (0,0) -- (8.93,4.29) -- cycle    ;

\draw (96,110) node   [align=left] {WAIT};
\draw (335,109) node   [align=left] {PURIF};
\draw (494,309) node   [align=left] {RELEASE};
\draw (175.5,259) node   [align=left] {ELIGIBLE};
\draw (150,24) node [anchor=north west][inner sep=0.75pt]  [font=\footnotesize,color={rgb, 255:red, 46; green, 128; blue, 226 }  ,opacity=1 ] [align=left] {Incoming SWAP\_UPDATE from \\a node with rank $\displaystyle r-1$.};
\draw (187,64) node [anchor=north west][inner sep=0.75pt]  [font=\footnotesize,color={rgb, 255:red, 208; green, 2; blue, 27 }  ,opacity=1 ] [align=left] {$\displaystyle i\ :=\ 0$};
\draw (576,109) node   [align=left] {PENDING};
\draw (371,40) node [anchor=north west][inner sep=0.75pt]  [font=\footnotesize,color={rgb, 255:red, 46; green, 128; blue, 226 }  ,opacity=1 ] [align=left] {Link (or segment) purification starts.};
\draw (418,132) node [anchor=north west][inner sep=0.75pt]  [font=\footnotesize,color={rgb, 255:red, 46; green, 128; blue, 226 }  ,opacity=1 ] [align=left] {purification\\succeeds.};
\draw (409,171) node [anchor=north west][inner sep=0.75pt]  [font=\footnotesize,color={rgb, 255:red, 208; green, 2; blue, 27 }  ,opacity=1 ] [align=left] {$\displaystyle i\ :=\ i\ +\ 1$};
\draw (519,196) node [anchor=north west][inner sep=0.75pt]  [font=\footnotesize,color={rgb, 255:red, 46; green, 128; blue, 226 }  ,opacity=1 ] [align=left] {purification\\fails.};
\draw (359,221) node [anchor=north west][inner sep=0.75pt]  [font=\footnotesize,color={rgb, 255:red, 46; green, 128; blue, 226 }  ,opacity=1 ] [align=left] {Endpoint is used as\\purification ancilla.};
\draw (187,161) node [anchor=north west][inner sep=0.75pt]  [font=\footnotesize,color={rgb, 255:red, 46; green, 128; blue, 226 }  ,opacity=1 ] [align=left] {$\displaystyle i\ ==\ P_{r}$};
\draw (268,298) node [anchor=north west][inner sep=0.75pt]  [font=\footnotesize,color={rgb, 255:red, 46; green, 128; blue, 226 }  ,opacity=1 ] [align=left] {Endpoint is swapped.\\(only on intermediate nodes)};
\draw (220,349) node [anchor=north west][inner sep=0.75pt]  [font=\footnotesize,color={rgb, 255:red, 208; green, 2; blue, 27 }  ,opacity=1 ] [align=left] {Save swap result.\\Send SWAP\_UPDATE if needed.};
\draw (32,177) node [anchor=north west][inner sep=0.75pt]  [font=\footnotesize] [align=left] {Endpoint arrives\\from link layer.};
\draw (502,349) node [anchor=north west][inner sep=0.75pt]  [font=\footnotesize] [align=left] {Endpoint is released\\to link layer.};
\draw (46,334) node [anchor=north west][inner sep=0.75pt]  [font=\footnotesize,color={rgb, 255:red, 144; green, 19; blue, 254 }  ,opacity=1 ] [align=left] {Deliver connection to the user\\(only on end nodes)};

\end{tikzpicture}
            }
    }
    \vspace{-.35cm}
    \caption{Finite state machine for an endpoint on a node with rank $r$. Blue labels indicate the conditions for the state transition, whereas red labels indicate the action performed during the transition.}
    \label{fig:states}
\end{figure}
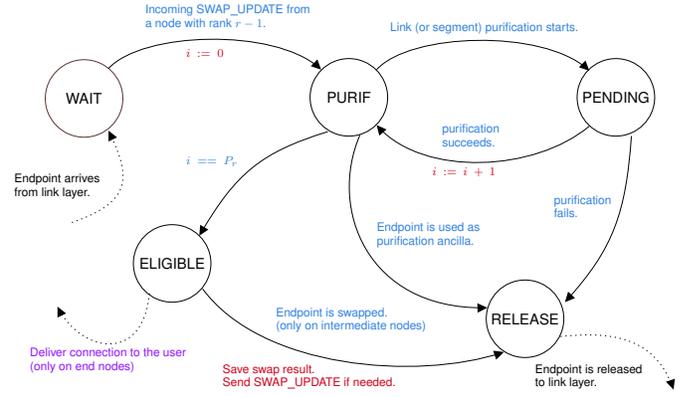

The generation of entangled connections begins when all nodes have computed their swapping neighbors and destinations. As shown in Fig. \ref{fig:states}, during this phase, an endpoint held by an $r$-ranked node of a connection can be in one of the following possible \emph{logical states}:
\begin{itemize}
    \item WAIT is the initial state for endpoints signaled by the lower layer when an entangled qubit pair is generated.
    \item PURIF is entered if  $r=0$, otherwise it is entered when a SWAP\_UPDATE message arrives from a node whose rank is $r-1$. An endpoint stays in this state if the link, segment, or connection must be purified.
    \item PENDING is entered when an endpoint is submitted to the purification procedure. If the procedure succeeds, the state transitions back to PURIF and the purification counter $i$ is increased by one.
    \item RELEASE is the transitional state of endpoints that are about to be released back to the Link layer. It is reached when the endpoint is used as an ancilla for purification, when the endpoint is swapped, or when purification fails.
    \item ELIGIBLE is the state for endpoints that are ready for entanglement swapping (on intermediate nodes), or that are ready to be delivered to the user (on end nodes). It is entered when the endpoint in PURIF state has been successfully purified a number of times equal to $P_r$.
\end{itemize}

\subsubsection{Swap Updates}
\label{subsubsec:updates}

SWAP\_UPDATE messages share some similarities with the TRACK message from \cite{Kozlowski_2020}, in the sense that they serve the purpose of keeping track of what endpoints have been swapped and the final Bell state of the new segment. The difference lies in the fact that TRACK messages are always forwarded by all nodes on the path towards an end node, whereas SWAP\_UPDATEs are forwarded by swapping neighbors and delivered to a swapping destination which, as highlighted before, may or may not be an end node.

\begin{figure} [tb]
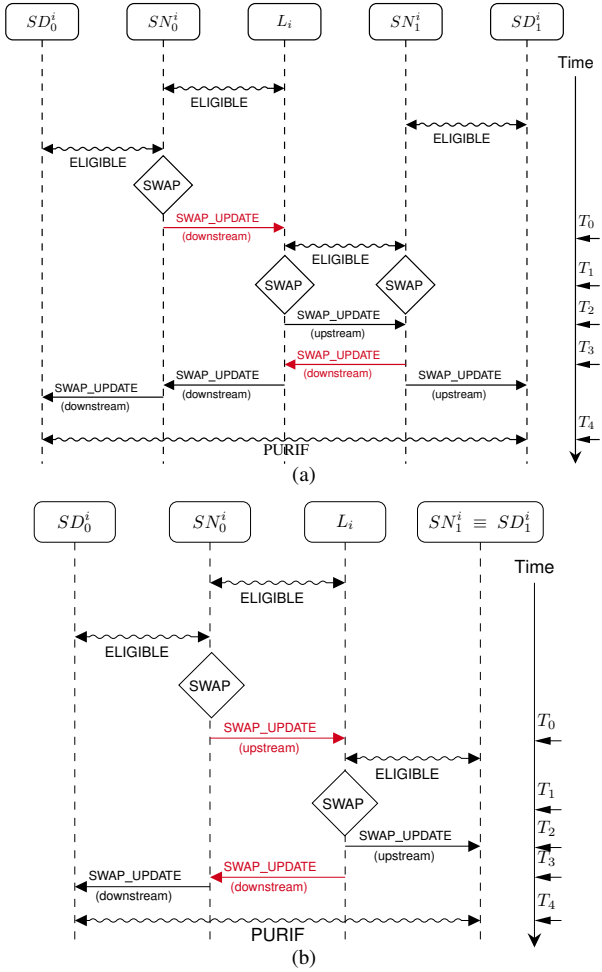

    \captionsetup[subfloat]{farskip=5pt,captionskip=0pt}
    \centering
    \subfloat[]{
        \tikzset{every picture/.style={line width=0.75pt}} 
        {\fontfamily{phv}\selectfont
            \resizebox{.9\linewidth}{!}{%

            }
        } \label{fig:seq_swap_co}
    }
    \caption{Example sequence of REDiP when (a) swapping neighbors and destinations are distinct and (b) upstream swapping neighbor and destination coincide. Waved arrows indicate that endpoints of an entangled link (or segment) between the two nodes have transitioned to a new state. The generation of a SWAP\_UPDATE message is highlighted in red.}
    \label{fig:seq_swap}
\end{figure}

Fig. \ref{fig:seq_swap_non} shows the behavior of an $r$-ranked node $L_i$ when its swapping neighbors and destinations do not coincide. We can see that $L_i$ receives an upstream SWAP\_UPDATE from  ${SN}_0^i$ (time $T_0$), but it has not swapped its two local endpoints yet, so the message cannot be immediately forwarded. As soon as $L_i$ has an upstream endpoint in ELIGIBLE state, it is swapped with the endpoint specified by the received message (time $T_1$). The final Bell state measurement is collected and saved in a temporary record as it is done in \cite{Kozlowski_2020}, and the two swapped endpoints transition to the RELEASE state. At this point $L_i$ can add the stored measurement result to the upstream SWAP\_UPDATE, which is forwarded to ${SN}_1^i$ (time $T_2$). When $L_i$ receives the SWAP\_UPDATE from  ${SN}_1^i$, it immediately updates and forwards downstream the message (time $T_3$). When both SWAP\_UPDATEs are delivered to the corresponding swapping destination, the two endpoints of the new segment between ${SD}_0^i$ and ${SD}_1^i$ transition to the state PURIF (time $T_4$). Fig. \ref{fig:seq_swap_co} shows the complementary case where the upstream swapping neighbor and destination coincide. The key difference with the previous example is that after the swap (time $T_1$), $L_i$ must generate the downstream SWAP\_UPDATE and send it to ${SN}_0^i$ (time $T_3$). Of course, in the mirror case of downstream coincidence (${SN}_0^i\ \equiv\ {SD}_0^i$), then $L_i$ would generate and send the upstream SWAP\_UPDATE instead of the downstream one.

\subsubsection{Purification Procedure}

\begin{figure} [tb]
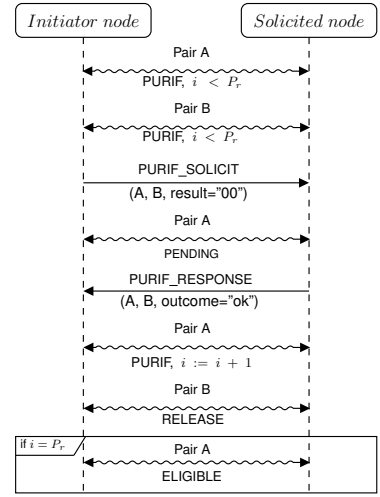

    \centering

    \tikzset{every picture/.style={line width=0.75pt}} 
    {\fontfamily{phv}\selectfont
        \resizebox{.55\linewidth}{!}{%

        }
    }
    \caption{Example sequence of REDiP purification. Waved arrows indicate that both endpoints of the link (or segment) have transitioned to a new state. For the PURIF state, there is also the new value of the purification counter $i$.}
    \label{fig:purif}
\end{figure}

What is still missing to generate connections is the link (or segment) purification procedure that takes endpoints from PURIF state and eventually turns them into ELIGIBLE. The procedure can be carried out between two swapping neighbors or between a node and one of its swapping destinations. In the latter case the endpoint residing on the swapping destination starts in WAIT state instead of PURIF and it never transitions to ELIGIBLE in case of successful purification. The two actors are called \emph{Initiator node} and \emph{Solicited node}, where the former has the initial task of choosing the link to purify and one or more other links as ancillas, depending on the purification algorithm. The policy used for this choice also depends on the chosen purification algorithm. For example, a \emph{recurrence} purification pattern like Deutsch algorithm \cite{Deutsch_1996} requires that ancillas must have the same fidelity as the purified link,  whereas an entanglement \emph{pumping} pattern like the one proposed in \cite{Dur_1999} allows the exploitation of ancillas with a lower fidelity than the purified link. The Initiator node then executes the purification algorithm and sends a PURIF\_SOLICIT message containing the identifier of the link to purify, ancillas and their purification measurement outcomes. The Solicited node receives the message and applies itself the purification algorithm. If the measurement outcome matches with the one in the PURIF\_SOLICIT message, then the purification was successful. In any case, the Solicited node replies with a PURIF\_RESPONSE message, containing the purified link identifier and ancillas (piggyback approach), and the purification outcome (OK or FAIL). If the purification was successful, the purification counter $i$ is increased by one, otherwise both endpoints are released back to the Link layer. Ancillas are always released to the Link layer right after their measurement. We show an example in Fig. \ref{fig:purif} where the purification requires only one ancilla (pair B). This is the case for popular purification patterns like Deutsch algorithm \cite{Deutsch_1996}. The procedure is repeated until the purification counter $i$ reaches the value $P_r$, where $r$ is the rank of the Initiator node. To determine the actor roles there is a simple rule: the lowest ranked node is the Initiator. The tie-breaking rule is left up to REDiP implementation; one could simply set the downstream node to always be the Initiator when the ranks are equal.

\subsection{Implementing Different Strategies}
\label{subsec:2d}
In this section we show how REDiP configurability allows to implement the Parallel ESS, Consecutive ESS, and Nested ESS. For each one of these strategies, we provide a way to compute the rank vector $R$ given the number of nodes $N$.

\subsubsection{Parallel strategy}
If we set the ranks as $R=[1,0,0,\ldots,0,0,1]$, we reproduce the Parallel ESS used in \cite{Kozlowski_2020}. However, this REDiP configuration performs slightly better because SWAP\_UPDATEs are generated by the nodes adjacent to the end node. TRACK messages from \cite{Kozlowski_2020} are instead generated by the end nodes themselves. This saves one transmission hop for each SWAP\_UPDATE, which translates in less idle time spent by qubits inside of the end nodes quantum memories.
\subsubsection{Consecutive strategy}
The rank vector $R=[N-2,0,1,2,\ldots,N-3,N-2]$ reproduces the Consecutive ESS from \cite{Li_2022}, where entanglement swapping is sequentially carried out by intermediate nodes along the path in the upstream direction. According to this REDiP version, it is not up to the downstream end node to trigger entanglement swapping on the successive repeater, which is instead the approach used in \cite{Li_2022}. This again saves one transmission hop for each SWAP\_UPDATE, leading to the same improvement seen for the Parallel ESS.
\subsubsection{Nested strategy}
The Nested ESS is implemented by a rank vector $R$ where each element $R_i$ is defined as follows:
\begin{equation}
    R_i = \max_{r \in \{0,1,\ldots,k\}} \{r \mid i \mod{2^r} = 0\},
\end{equation}
where $N=2^k+1$ for some $k>0$. This last condition is an assumption for the applicability of this strategy.

\section{Simulation Analysis}
\label{sec:3}
We implemented a REDiP simulator to estimate the impact of several PESSs on the protocol performances in terms of throughput (\textit{pairs}/\textit{s}) and fidelity. The simulator was implemented as a Python package on top of Netsquid \cite{Coopmans_2021}, a simulation engine for quantum networks. Below we define the assumptions of the simulation campaign and then we show some representative scenarios. From the simulation outcomes, it is clear that the assignments of REDiP ranks and purification rounds have a significant impact on the entanglement distribution performance.   

\subsection{Assumptions}
\label{sec:3a}
In our simulation scenarios we assume all repeaters and links have the same hardware specifications. In particular, we modeled an implementation of the Midpoint Source (MS for short) Link layer protocol defined in \cite{Jones_2016}, and we set its parameters according to the \emph{optimistic} parameter set used in \cite{Yoshida_2020}. The rationale behind this choice is that \cite{Yoshida_2020} employs the most advanced emerging technology for multi-mode quantum memories.

To model quantum errors, we introduced three sources of noise: (i) An initial depolarization probability $p^d_0$ applied on both endpoints of a newly generated link, so that its initial fidelity is $F_0<1$. (ii) A quantum memory exponential dephasing rate $r_d$, which simulates the decoherence of a qubit state $\rho$ stored inside a quantum memory, so that after $\Delta t$ time the new state $\rho'$ of the qubit can be written as
\begin{equation}
    \rho' = f(\rho, \Delta t) = e^{-r_d \Delta t}\rho + (1 - e^{-r_d \Delta t}) Z\rho Z.
\end{equation}
We also define the coherence time $T_c$ of a quantum memory as the time after which the dephasing probability reaches $5\%$. 
(iii) A depolarization probability $p_{err}^d$ applied right before each measurement and quantum gate, to introduce a noise component in entanglement swapping and purification operations.

In our simulations we considered a network topology composed of 9 nodes connected as a linear chain where all fiber-optic links are $15km$ long. We used Deutsch algorithm \cite{Deutsch_1996} whenever entanglement purification was required. We set $p_0^d=0.01334$ so that the initial fidelity of link-generated pairs is approximately $F_0=0.98$.

\subsection{Results}
\label{subsec:3b}

\begin{figure*} [t]
    \captionsetup[subfloat]{farskip=0pt,captionskip=0pt}
    \centering
    \subfloat[]{\includegraphics[width=.45\linewidth]{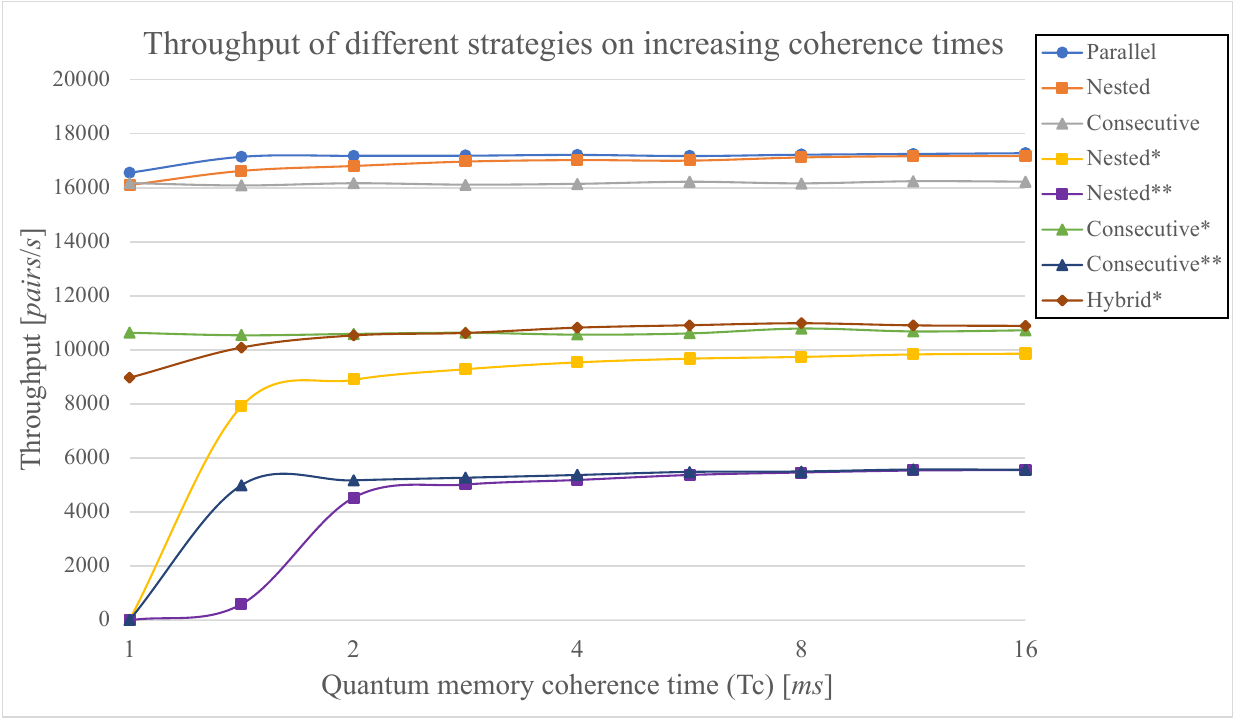}\label{fig:results_a}}
    \hspace{.2cm}
    \subfloat[]{\includegraphics[width=.45\linewidth]{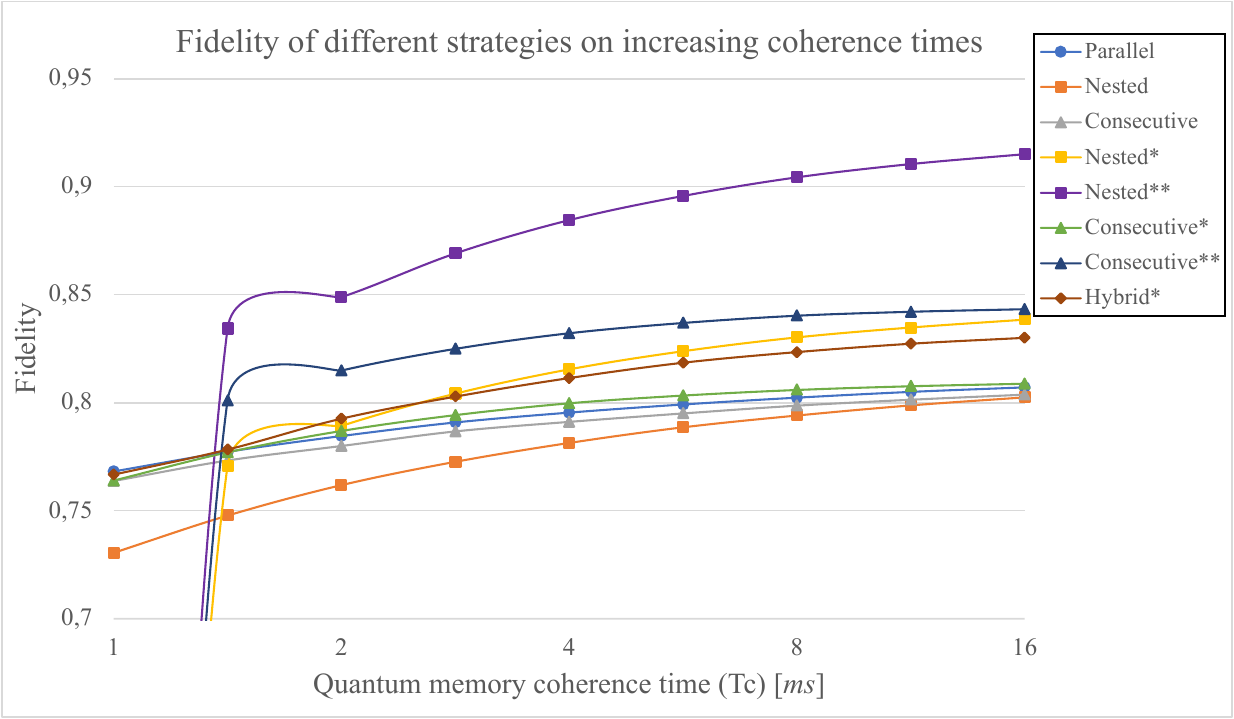}\label{fig:results_b}}
    \hfill
    \subfloat[]{\includegraphics[width=.45\linewidth]{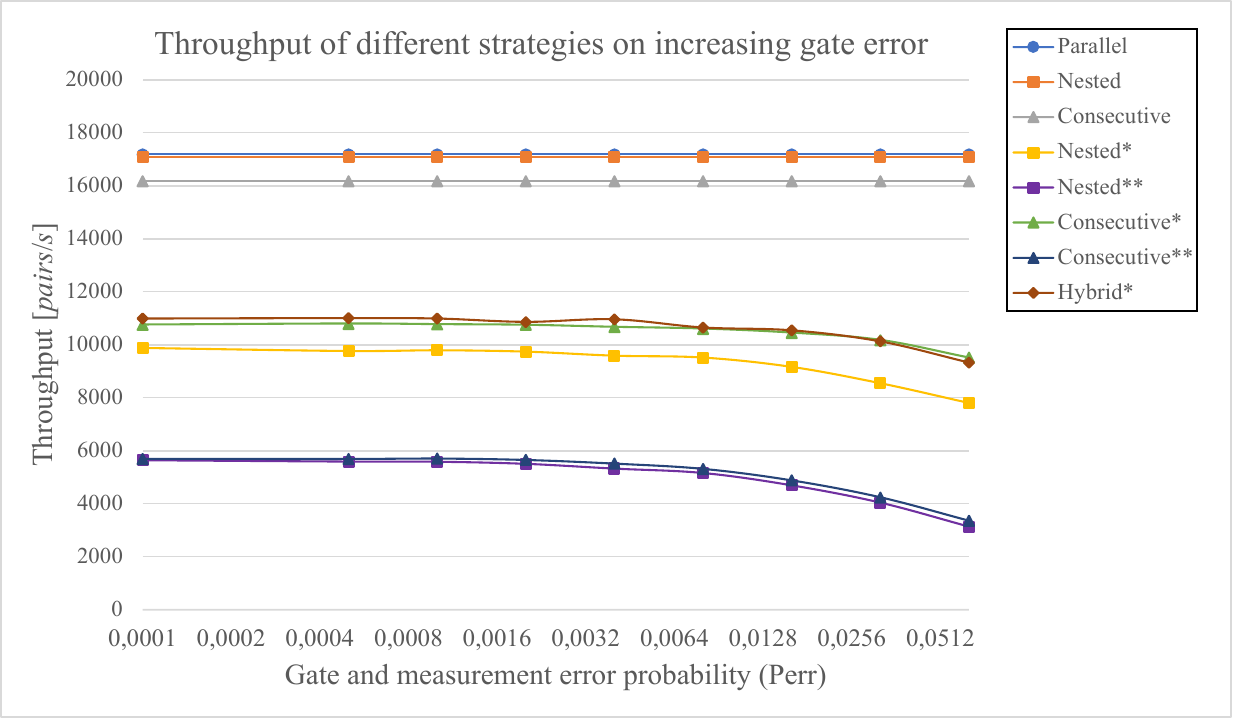}\label{fig:results_c}}
    \hspace{.2cm}
    \subfloat[]{\includegraphics[width=.45\linewidth]{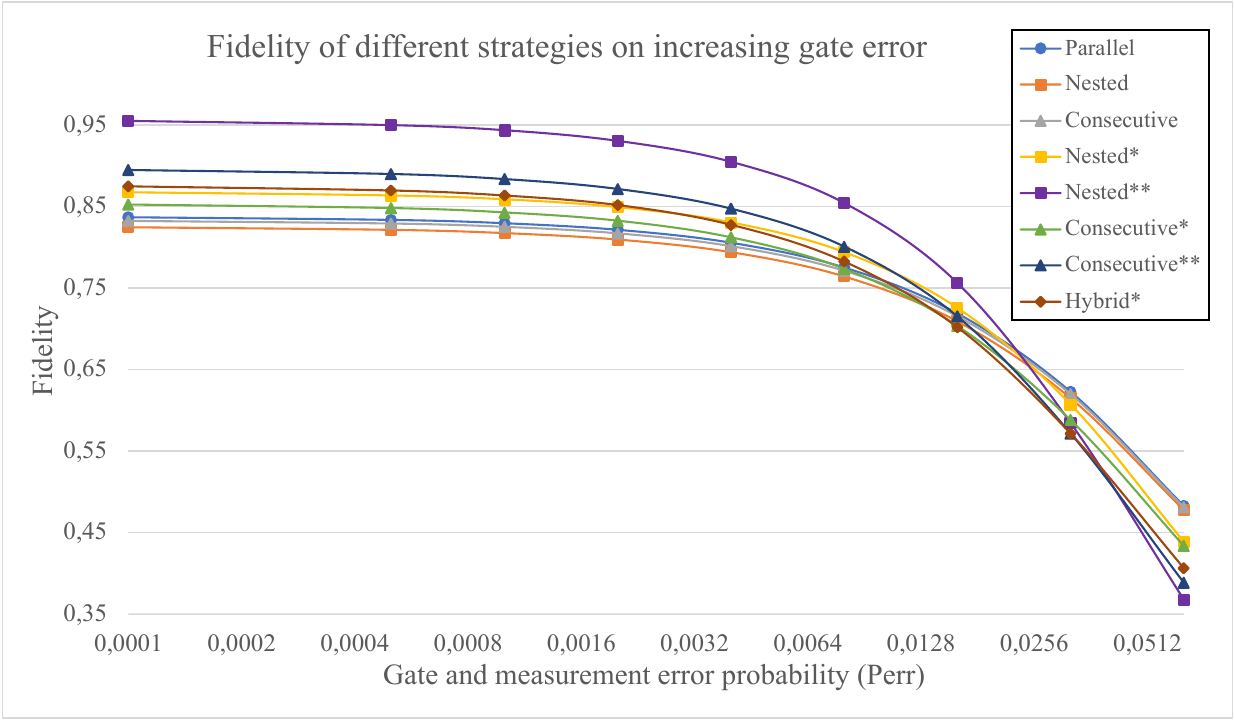}\label{figresults_d}}
    \caption{Results of the simulation campaign on a nine nodes chain, showing (a) throughput, (b) fidelity as a function of memory coherence time $T_c$, and (c) throughput, (d) fidelity as a function of gate error $p^d_{err}$. Each marker was obtained as the average of 31 independent runs. Confidence intervals are not shown as their size is smaller or comparable to the markers.}
    \label{fig:results}
\end{figure*}

Fig. \ref{fig:results} reports the results of our simulations. Specifically, Figs. \ref{fig:results_a} and \ref{fig:results_b} plot the end-to-end throughput (\textit{pairs/s}) and average fidelity as a function of coherence time $T_c$, keeping $p^d_{err}=0.005$ as a constant, whereas Figs. \ref{fig:results_c} and \ref{figresults_d} plot the same metrics as a function of $p^d_{err}$, keeping $T_c = 5ms$ as a constant. In all these plots, we compare the performances of a set of eight representative PESSs. For the sake of readability, we named each PESS by the name of the entanglement swapping strategy and postponed a number of stars ("$*$") equal to the total rounds of purification applied. To avoid ambiguities, we provide in Table \ref{tab:table1} the vectors $R$ and $P$ used on each PESS (we do not repeat $R$ across strategies sharing the same ESS). Regarding the \emph{Hybrid*} PESS, it has been added to highlight how REDiP flexibility allows to mix up different strategies: $0$-ranked nodes are assigned as in the Nested ESS, but all remaining repeaters are $1$-ranked, so that they swap in parallel. This behavior makes this strategy a hybrid between Nested and Parallel.

We can see from Figs. \ref{fig:results_a} and \ref{fig:results_c} that the throughput highly depends on the total number of purification rounds, and that purification performed at higher ranks costs slightly more in terms of throughput due to the longer transmission time of purification control messages, which is why \emph{Nested*} has a lower throughput than \emph{Consecutive*} and \emph{Hybrid*}. For what concerns fidelity, Figs. \ref{fig:results_b} and \ref{figresults_d} clearly show that \emph{Nested**} outperforms all other PESSs involved in these simulations. This makes it particularly suitable for stringent fidelity requirements, at the cost of a lower throughput. As a general consideration emerging from these results, we can say that purification combined with the Consecutive ESS, given its intrinsic asymmetry, does not provide good performances with respect to other strategies. Moreover, REDiP can be setup by the user with a suitable PESS that aims to maximize the throughput while delivering entanglement with a guaranteed minimum fidelity. For example, \emph{Parallel} is a good option for low fidelity requirements, especially when $T_c$ is critically low. Finally, we see from Fig. \ref{fig:results_a} that some strategies with one and two degrees of purification are not able to deliver entanglement if the coherence time $T_c$ is below a certain threshold. This happens because qubits are automatically released by a \emph{cutoff} mechanism if they stay in a quantum memory for a time higher than $T_c$.


\begin{table}[tb]
    \small
    \centering
    \caption{{PESS SPECIFICATIONS}}
    \label{tab:table1}
    \begin{tabular}{|l|l|l|}
        \hline
        PESS & $R$ & $P$\\
        \hline
        \emph{Parallel} & $[1,0,\ldots,0,1]$ & $[0,0]$ \\
        \hline
        \emph{Nested} & $[3,0,1,0,2,0,1,0,3]$ & $[0,0]$ \\
        \hline
        \emph{Nested*} &" & $[0,0,1,0]$ \\
        \hline
        \emph{Nested**} & " & $[0,1,1,0]$ \\
        \hline
        \emph{Consecutive} & $[7,0,1,2,3,4,5,6,7]$ & $[0,\ldots,0]$ \\
        \hline
        \emph{Consecutive*} & " & $[0,0,1,0,\ldots,0]$ \\
        \hline
        \emph{Consecutive**} & " & $[0,0,1,0,1,0,0,0]$ \\
        \hline
        \emph{Hybrid*} & $[2,0,1,0,1,0,1,0,2]$ & $[0,1,0]$ \\
        \hline
    \end{tabular}
\end{table}

\section{Conclusion}
\label{sec:4}
In this paper we presented REDiP, a quantum data plane protocol for entanglement distribution that integrates both entanglement swapping and purification to provide a configurable service to the upper layer. A preliminary performance evaluation of REDiP proved that its configurability is essential to meet user requirements and mitigate the effect of quantum errors. We left as further study the optimization problem to assign ranks and purification rounds of a REDiP tunnel, given a set of user requirements and network conditions.

\bibliographystyle{ieeetr}
\bibliography{refs}

\end{document}